\documentclass[
preprint,
amsmath,
amssymb,
aps,
prd,
longbibliography
]{revtex4-2}
\usepackage{booktabs}
\usepackage{graphicx}
\usepackage{dcolumn}
\usepackage{bm}
\usepackage[bookmarks=true,colorlinks=true,linkcolor=blue,unicode=true]{hyperref}
\usepackage[mathlines]{lineno}
\usepackage{xcolor} 
\usepackage{color} 

\newcounter{YJC}

\begin{document}

\title{Expressibility and trainability of a two-dimensional pairwise quantum-circuit ansatz}

\author{Shuai Zhang}
\author{Wei Liu}
\author{Ji-Chong Yang}
\email{yangjichong@lnnu.edu.cn}
\thanks{Corresponding author}

\affiliation{Department of Physics, Liaoning Normal University, Dalian 116029, China}
\affiliation{Center for Theoretical and Experimental High Energy Physics, Liaoning Normal University, Dalian 116029, China}

\date{\today}

\begin{abstract}
Parameterized quantum circuits~(PQCs) constitute a central building block of variational quantum algorithms~(VQAs) and quantum machine learning~(QML) methods. 
Existing ansatz designs often adopt hardware-agnostic or simplified 1D chain/ring entanglement patterns. 
However, as quantum hardware continues to develop, native 2D connectivity patterns, such as planar superconducting-qubit architectures, are becoming increasingly important. 
Inspired by this hardware structure, we construct a native 2D pairwise ansatz and compare its expressibility and trainability with representative 1D ansatze at identical layer depths, despite their different circuit depths.
For the fixed 16-qubit system, the 2D ansatz has the smallest KL divergence at $L=1$ and $2$, and its second-order frame potential approaches the theoretical lower bound more rapidly at shallow layer counts than the frame potentials of the three 1D ansatze.
We also evaluate the gradient variance of the Pauli-$Z$-string expectation value $\langle Z_0\otimes\cdots\otimes Z_{15}\rangle$ with respect to the first $R_y$ angle.
For this Pauli-$Z$ string and fixed parameter, the gradient variance is smaller for the 2D circuit at $L=1$--$4$. 
The differences narrow at $L=5$, and the four ansatze yield statistically compatible variances at $L=6$.
\end{abstract}

\maketitle

\section{\label{sec:1}Introduction}
At present, 1D ansatz have become relatively mature and commonly adopt nearest-neighbor chain  structures. 
Their advantages include shallow circuit depth, a small number of gates, good compatibility with noisy intermediate-scale quantum~(NISQ) devices~\cite{Arute:2019zxq,Preskill:2018jim}, and straightforward mapping onto linear or near-linear coupling hardware.
More generally, parameterized quantum circuits~(PQCs) constitute a central building block of variational quantum algorithms, such as the variational quantum eigensolver~(VQE)~\cite{Kandala:2017vok,Peruzzo:2013bzg,McClean:2015vup,OMalley:2016ugf,Cerezo:2020jpv,Tilly:2021jem}, the quantum approximate optimization algorithm~(QAOA)~\cite{Farhi:2014ych}, variational quantum searching neighbor~(VQSN)~\cite{Yang:2024bqw}, the quantum variational
error corrector (QVECTOR)~\cite{Johnson:2017tti}, and quantum neural networks (QNN)~\cite{farhi2018,Havlicek:2018nqz,Schuld:2018ahn}, among others. 
They are particularly relevant to recent superconducting quantum processors, where hardware-efficient ansatz have been experimentally demonstrated~\cite{OMalley:2016ugf,Kandala:2017vok}.
However, the limitations of 1D ansatze are also evident. 
Entanglement propagation is slow, and establishing correlations between distant qubits typically requires multiple circuit layers.
As a result, their expressibility is relatively limited. 
In particular, when dealing with some 2D lattice models, 1D circuits can require a significant increase in depth, which further leads to noise accumulation and exacerbates trainability issues, such as the barren plateaus~\cite{McClean2018,Ragone_2024}.

Compared to ansatze based on 1D chain or ring, native 2D pairwise ansatz offers several advantages.
This design is also naturally motivated by superconducting quantum processors, where qubits are commonly arranged in planar 2D layouts with local two-qubit couplings, as exemplified by programmable superconducting processors such as Sycamore and architectures based on the heavy hex topology~\cite{Arute:2019zxq,Kjaergaard:2019lmy,Krantz:2019jkw,Chamberland:2019zev}. 
On the one hand, a natively 2D entangling pattern better conforms to the physical connectivity of planar superconducting hardware, thereby reducing routing overhead and the need for additional SWAP gates. 
On the other hand, since the connectivity of an ansatz can strongly affect its expressibility and entangling capability~\cite{Sim_2019}, increasing the local connectivity of the entanglement from 1D to native 2D can improve the expressibility of shallow circuits.
Therefore, in this work, we design a native 2D pairwise entangled ansatz and evaluate its expressibility and trainability employing three complementary metrics, namely the Kullback--Leibler~(KL) divergence and the second-order~($t=2$) frame potential for expressibility, together with the gradient variance for trainability~\cite{Sim_2019,Brandao2016,McClean2018,Renes_2004,Dankert_2009}.
The primary comparison is performed at the same number of repeated rotation--entanglement blocks. 
This is a layer-matched, rather than resource-matched, comparison, i.e., a 2D block contains $24$ CNOT gates, whereas each 1D block contains $16$. 
Because no single resource normalization can simultaneously match gate count, compiled depth, routing overhead, and native connectivity without altering the ansatz definitions, we adopt the number of repeated ansatz layers as a transparent common basis for comparing their architecture-dependent behavior.

The remainder of this paper is structured as follows. 
Sec.~\ref{sec:2} is a brief introduction to the four ansatze. 
Sect.~\ref{sec:3} discusses the three evaluation metrics and their implementation. 
The numerical results of evaluation metrics are presented in Sec.~\ref{sec:4}.  
Sec.~\ref{sec:5} summarizes the main conclusions.

\section{\label{sec:2} The  structures of four ansatze}

The PQC is defined as a tunable unitary transformation $U(\boldsymbol{\theta})$ acting on a $N$-qubit reference state $\lvert \phi_{0} \rangle$, which is commonly chosen as $\lvert 0 \rangle^{\otimes N}$. 
The output quantum state generated by the circuit is therefore given by
\begin{equation}
\lvert \psi(\boldsymbol{\theta}) \rangle = U(\boldsymbol{\theta}) \lvert \phi_0 \rangle ,
\tag{1}
\label{eq.2.1}
\end{equation}
where $\boldsymbol{\theta}$ denotes a parameter vector, each drawn uniformly from $[0, 2\pi)$.
In this work, the parameters are implemented as rotation angles of parameterized quantum gates, such as the angle $\theta$ in $R_y(\theta)$. 
Adjusting these parameters changes the quantum operation and thus alters the evolution of the quantum state.

This work investigates four ansatze, three of which are 1D  structures based on the \texttt{Qiskit} \texttt{TwoLocal} framework~\cite{qiskit2024}. 
These ansatze include Circular, Pairwise, and Shifted Circular Alternating~(SCA).
All four ansatze operate on $16$ qubits, with each layer consisting of an $S$-layer, corresponding to a single-qubit rotation layer, followed by an entanglement layer. 
Finally, an additional $S$-layer is appended after all $L$ layers. 
The rotation layer consists of single-qubit gates that act on all qubits, while the entanglement layer uses two-qubit gates to establish quantum correlations between qubits.
In this work, $R_{y}$, $R_{z}$ and CNOT are chosen as rotation and entangling gates, respectively. 
Consequently, the total number of trainable parameters is $2N(L+1)$, where $N$ represents the number of qubits, and $L$ denotes the depth of the layers.

\subsection{\label{sec:2.1}1D pairwise} 

\begin{figure}[htbp]
\begin{center}
\includegraphics[width=0.8\hsize]{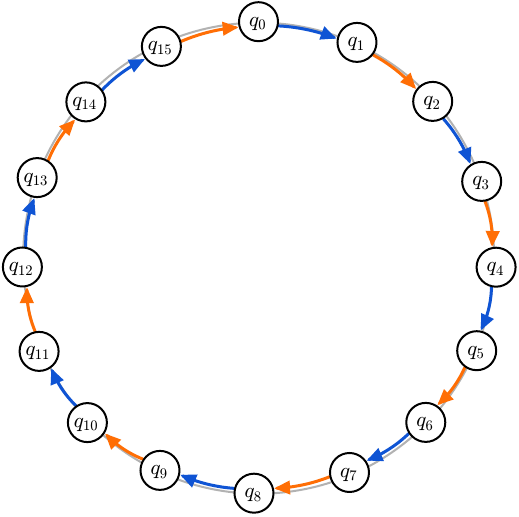}
\caption{\label{fig:1dpairwise}Connection diagram for a single layer of 16 qubits in a 1D pairwise ansatz. They are connected end-to-end to form a ring structure, where the blue arrows indicate the even-layer entanglement connections and the yellow arrows indicate the odd-layer entanglement connections.}
\end{center}
\end{figure}

\begin{figure}[htbp]
\begin{center}
\includegraphics[width=0.99\hsize]{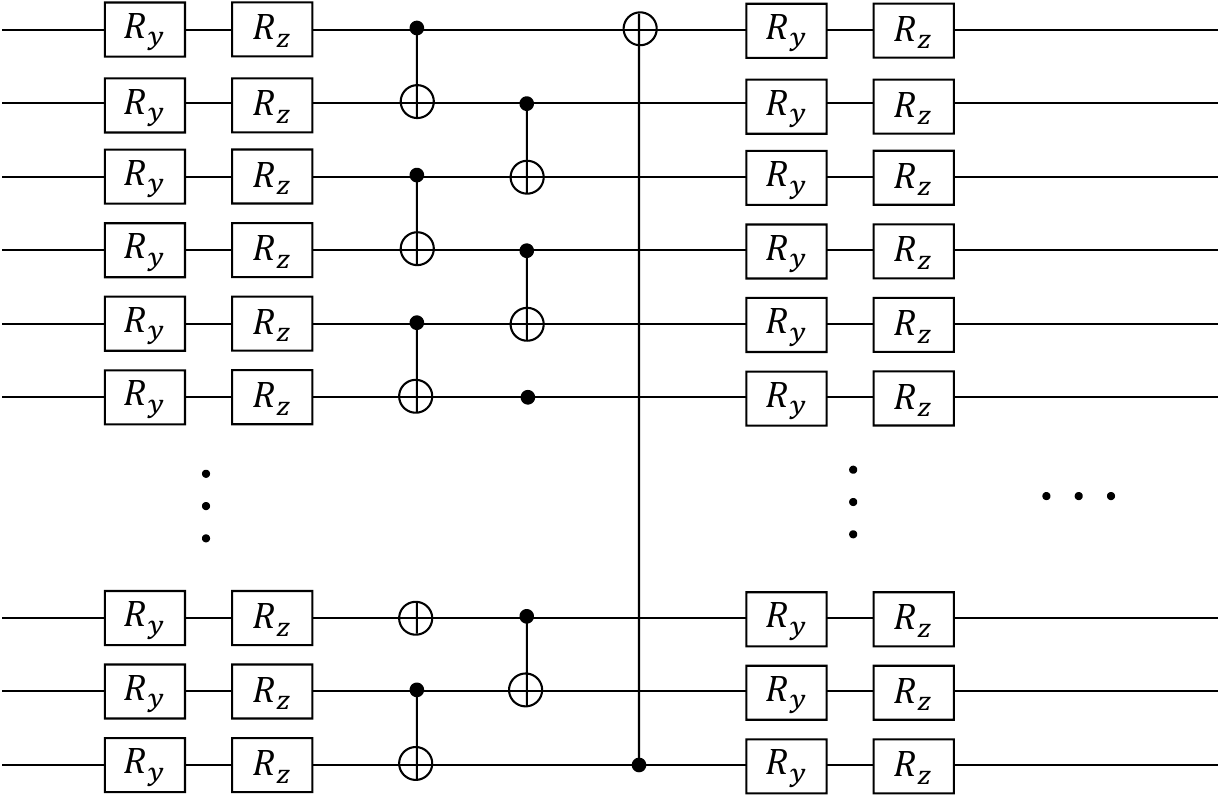}
\caption{\label{fig:circ1}The ansatz consists of 16 qubits arranged in 1D pairwise architecture. Each layer is composed of $R_y$ and $R_z$ rotation gates followed by CNOT entangling gates, and the same structure is repeated throughout all subsequent layers.}
\end{center}
\end{figure}
In this study, we construct a pairwise entanglement ansatz based on a 1D ring topology. 
The ansatz acts on $16$ qubits arranged as a closed 1D chain, namely $0-1-2-\cdots-15-0$. 
Each layer consists of a single-qubit rotation layer, referred to as the $S$-layer, followed by an entanglement layer. 
The entanglement layer adopts an even-odd staggered structure, which can be viewed as a local entanglement pattern similar to 1D brickwork circuit.

Specifically, the entanglement layer is further divided into an even-layer and an odd-layer. 
In the even-layer, qubits with even indices are used as control qubits and apply CNOT gates to their neighbors on the right, e.g., $0 \rightarrow 1, 2 \rightarrow 3,\ldots,14 \rightarrow 15$. 
In the odd-layer, qubits with odd indices are used as control qubits and apply CNOT gates to their neighbors on the right, e.g., $1 \rightarrow 2, 3 \rightarrow 4,\ldots,15 \rightarrow 0$. 
Therefore, each entanglement layer contains $16$ CNOT gates. 

This structure preserves local nearest-neighbor entanglement in 1D while covering the entire ring through the staggered operation pattern, and thus serves as 1D baseline model for the subsequent 2D topology-based ansatz.
The entanglement topology and the quantum circuit implementation of the 1D pairwise ansatz are shown in Fig.~\ref{fig:1dpairwise} and Fig.~\ref{fig:circ1}, respectively.

\subsection{\label{sec:2.2}1D circular } 

\begin{figure}[htbp]
\begin{center}
\includegraphics[width=0.8\hsize]{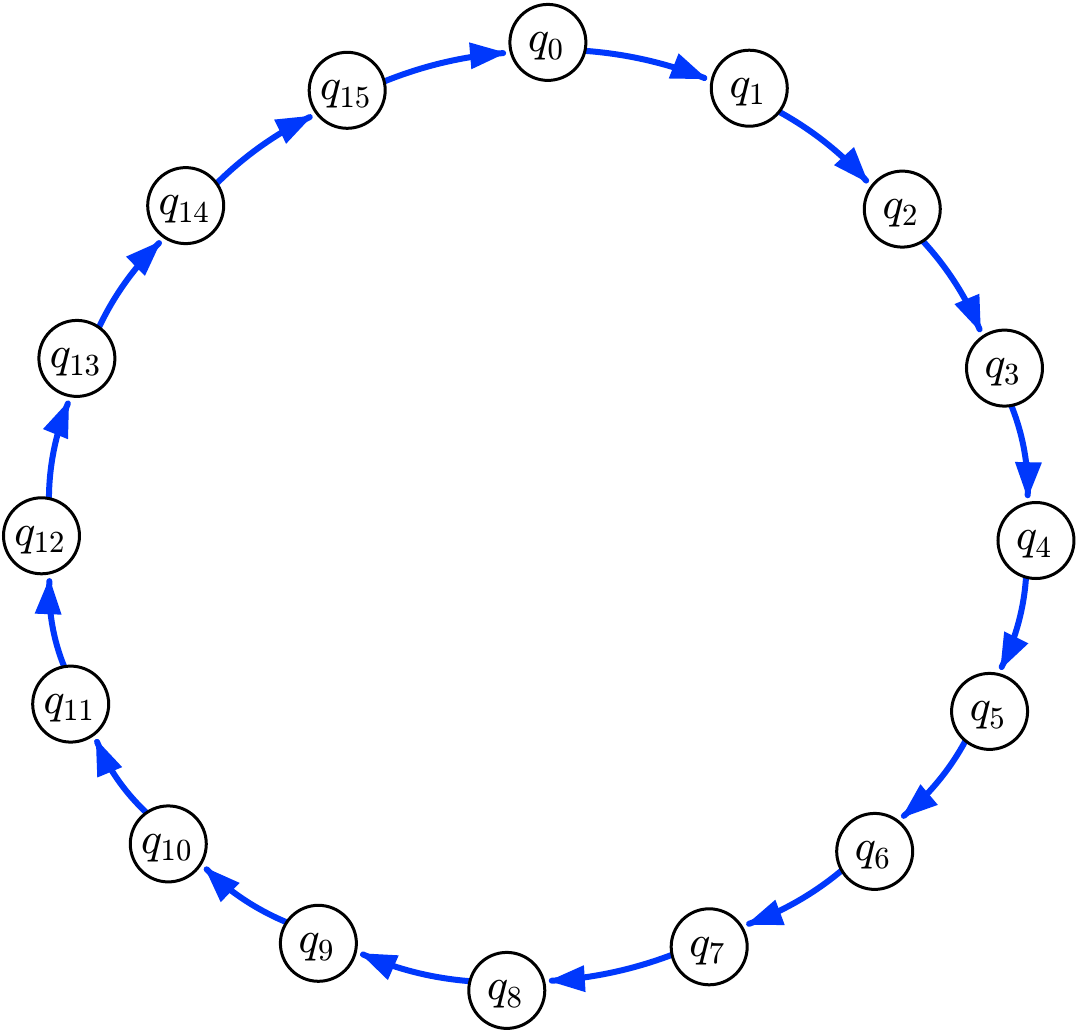}
\caption{\label{fig:1dcircular}Connection diagram for a single layer of 16 qubits in 1D circular ansatz. The same as  Fig.~\ref{fig:1dpairwise}, the arrows represent CNOT gates.}
\end{center}
\end{figure}

\begin{figure}[htbp]
\begin{center}
\includegraphics[width=0.99\hsize]{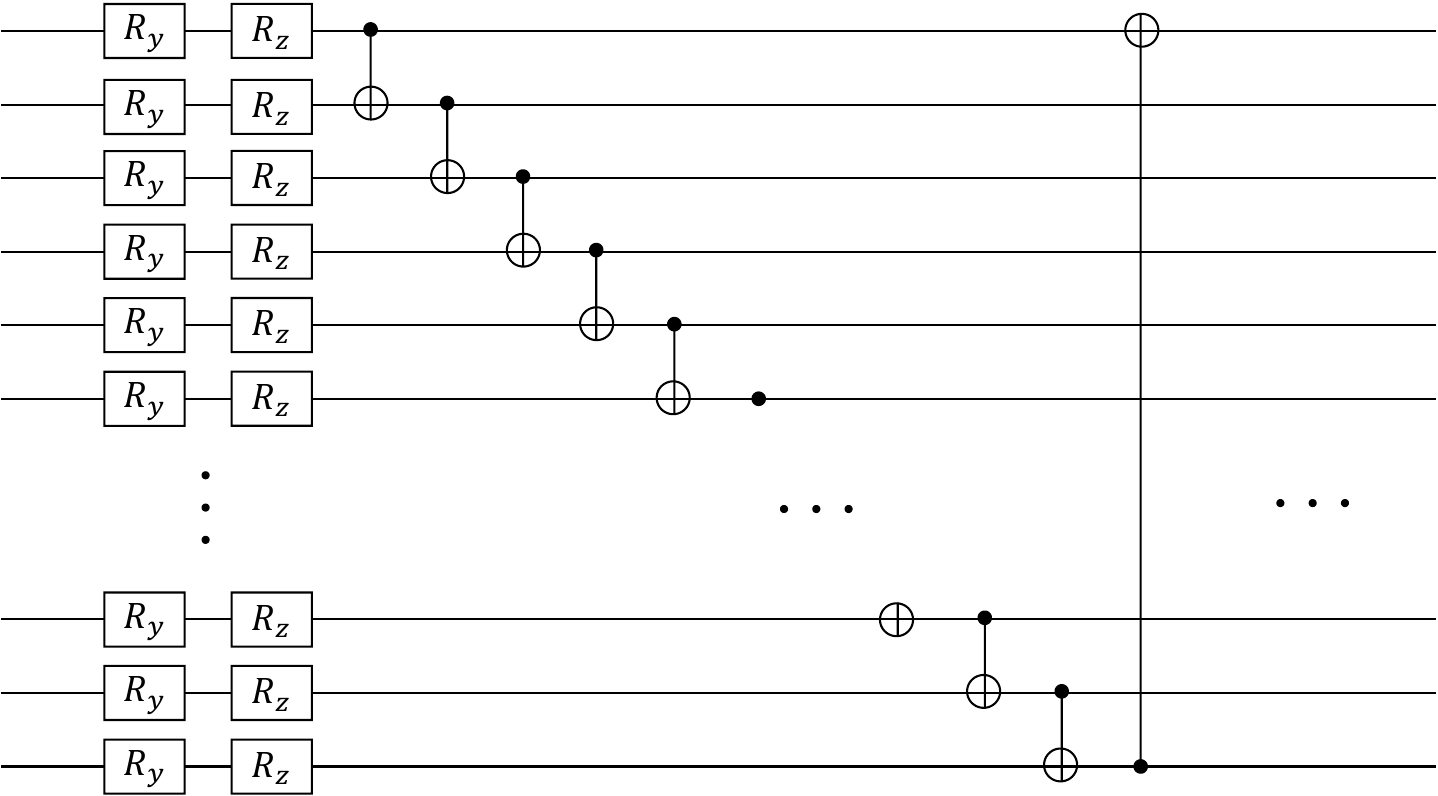}
\caption{\label{fig:circ2}The quantum circuit of a 1D circular consisting of 16 qubits. The structure of each layer is the same as Fig.~\ref{fig:circ1}, consisting of $R_y$ and $R_z$ rotation gates and CNOT gates.}
\end{center}
\end{figure}
1D circular ansatz adopts a 1D ring-shaped entanglement topology.
In each layer, a single-qubit rotation layer $S$ is followed by a complete circular CNOT chain, with entangling connections $0 \rightarrow 1, 1 \rightarrow 2, \ldots, 14 \rightarrow 15, 15 \rightarrow 0$. 
Thus, for $16$ qubits, each layer contains $16$ CNOT gates. 
Compared with linear entanglement, the circular structure introduces an additional entangling connection between the first and last qubits, thereby mitigating boundary effects and enabling a more uniform distribution of entanglement across the circuit. 
This design provides stronger global information propagation than a simple linear chain~\cite{Sim_2019}.
The entanglement topology and the quantum circuit implementation of the 1D circular ansatz are shown in Fig.~\ref{fig:1dcircular} and Fig.~\ref{fig:circ2}, respectively.

\subsection{\label{sec:2.3}SCA} 

\begin{figure*}[htbp]
\begin{center}
\includegraphics[width=0.99\hsize]{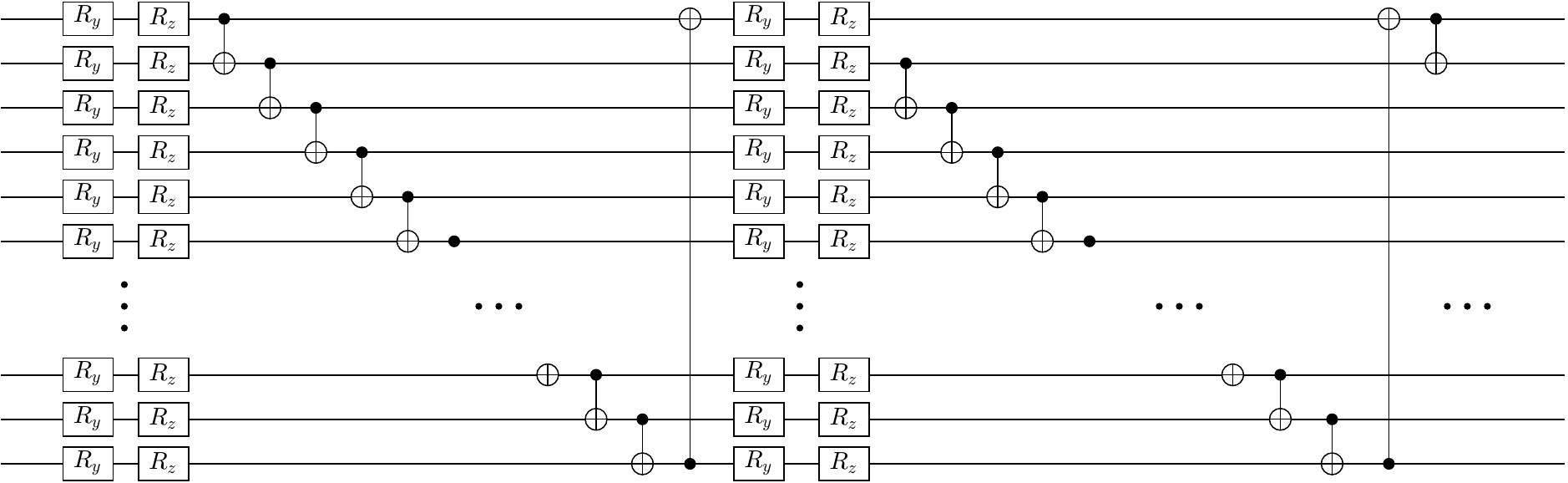}
\caption{\label{fig:circ3}The first two layers of a 16-qubit 1D SCA quantum circuit. Similar to Fig.~\ref{fig:circ1}, the circuit is composed of rotation and entangling layers, with the starting index of the entangling pattern shifted by one qubit between adjacent layers.}
\end{center}
\end{figure*}

SCA entanglement topology is a parameterized quantum circuit entanglement scheme designed for 1D ring-structured qubit systems. 
The basic idea is to apply a complete circular CNOT chain after the single-qubit rotation $S$-layer, so that all neighboring qubits are entangled according to a ring topology. 

For a system with $n=16$ qubits, each layer contains $16$ CNOT gates, covering the closed connectivity pattern $0\rightarrow1,1\rightarrow2,\ldots,15\rightarrow0$.
Unlike the circular entanglement pattern with a fixed starting point, the SCA ansatz shifts the starting index of the circular CNOT chain by one qubit between consecutive entangling layers. 
For the $l$-th entanglement block with $n=16$, the CNOT pattern can be written as
\begin{equation}
q_{(l+k)\bmod 16}
\rightarrow
q_{(l+k+1)\bmod 16},
\qquad
k=0,\ldots,15,
\tag{2}
\label{eq.2.2}
\end{equation}
where $l$ denotes the index of the entanglement block. 

This construction preserves the same circular connection and fixed CNOT direction in each block, while causing the order of entanglement operations between layers to shift periodically.
As a result, the SCA ansatz avoids repeating an identical gate sequence in every entangling layer while preserving the hardware-friendly 1D ring topology. 
The quantum circuit implementation of the 1D SCA ansatz is shown in Fig.~\ref{fig:circ3}.

\subsection{\label{sec:2.4}2D pairwise} 

The 2D pairwise entanglement topology is a hardware-efficient entangling scheme designed specifically for parameterized quantum circuits based on a 2D nearest-neighbor architecture.
In this structure, 16 qubits are fixed on a $4\times4$ 2D grid. 
Each qubit is labeled by its grid coordinate $(i,j)$, where $i,j\in\{0,1,2,3\}$. 
CNOT gates are restricted to nearest-neighbor interactions on this 2D grid.
The qubits are indexed in row-major order. 
For the $k$-th qubit located at coordinate $(i,j)$ on the 2D grid, its parity is defined as
\begin{equation}
p(k) = (i+j)\bmod 2,
\tag{4}
\label{eq.2.4}
\end{equation}
where qubits with $p(k)=0$ are called even-parity qubits; those
with $p(k)=1$ are odd-parity qubits. 
This parity decomposition allows nearest-neighbor interactions to be organized into direction-resolved CNOT sublayers while preserving the locality constraints of the underlying 2D topology.

\begin{figure}[htbp]
\begin{center}
\includegraphics[width=0.99\hsize]{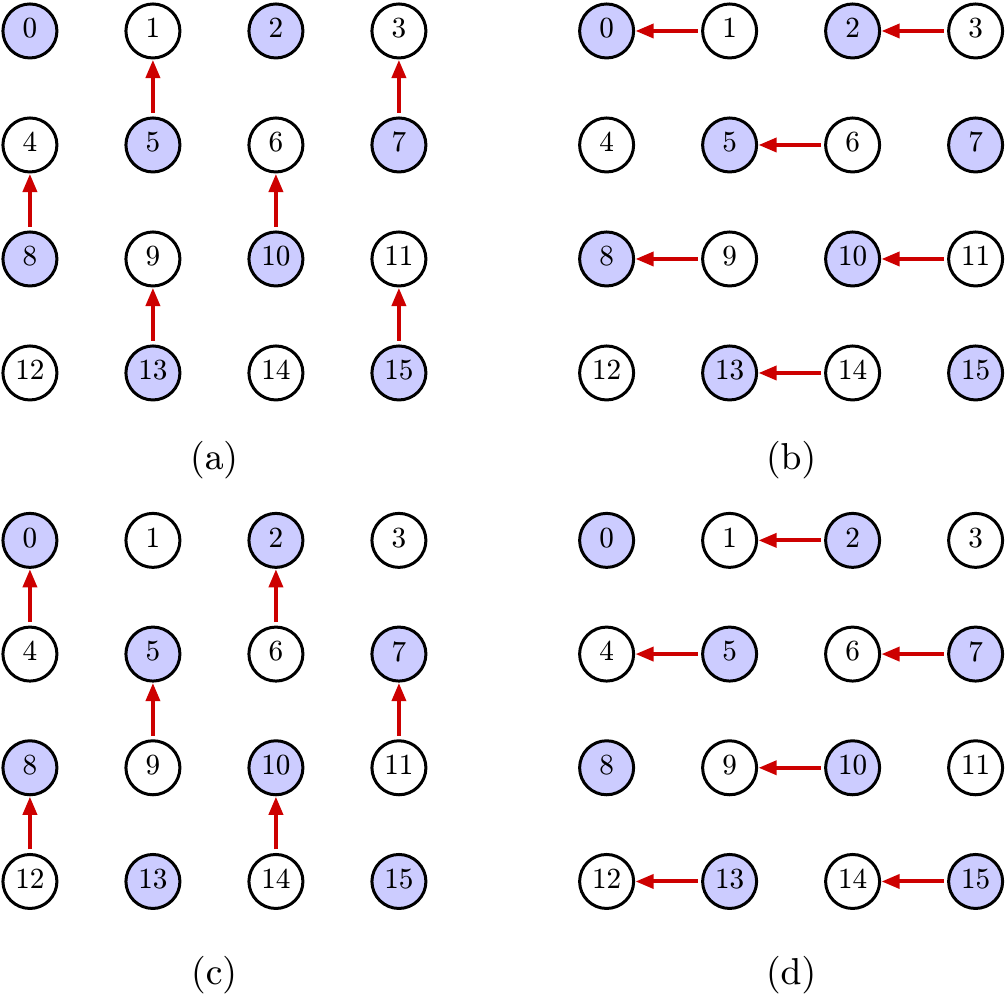}
\caption{\label{fig:pair2D}Four CNOT sublayers on a $4\times4$ grid ($N=16$). Panels (a) and (c) contain complementary sets of upward CNOT gates, while panels (b) and (d) contain complementary sets of leftward CNOT gates. 
The four sublayers in panels (a)--(d) are applied sequentially to form a complete entangling layer, corresponding to $U_1$, $L_1$, $U_2$, and $L_2$. 
Purple circles denote even-parity qubits, white circles denote odd-parity qubits, and each arrow points from the control qubit to the target qubit.}
\end{center}
\end{figure}

Each layer of the 2D pairwise ansatz consists of a single-qubit rotation $S$-layer , followed by four directional CNOT sublayers:
\begin{equation}
S \rightarrow U_1 \rightarrow L_1 \rightarrow U_2 \rightarrow L_2,
\tag{5}
\label{eq.2.5}
\end{equation}
where $U_1$ and $U_2$ contain complementary sets of upward CNOT gates and $L_1$ and $L_2$ contain complementary sets of leftward CNOT gates. 
Each sublayer contains six non-overlapping CNOT gates, so one complete entangling layer contains 24 CNOT gates and covers every horizontal and vertical nearest-neighbor edge of the $4\times4$ grid exactly once. 
Figure~\ref{fig:pair2D} shows these four sublayers.

\begin{figure*}[htbp]
\begin{center}
\includegraphics[width=0.85\hsize]{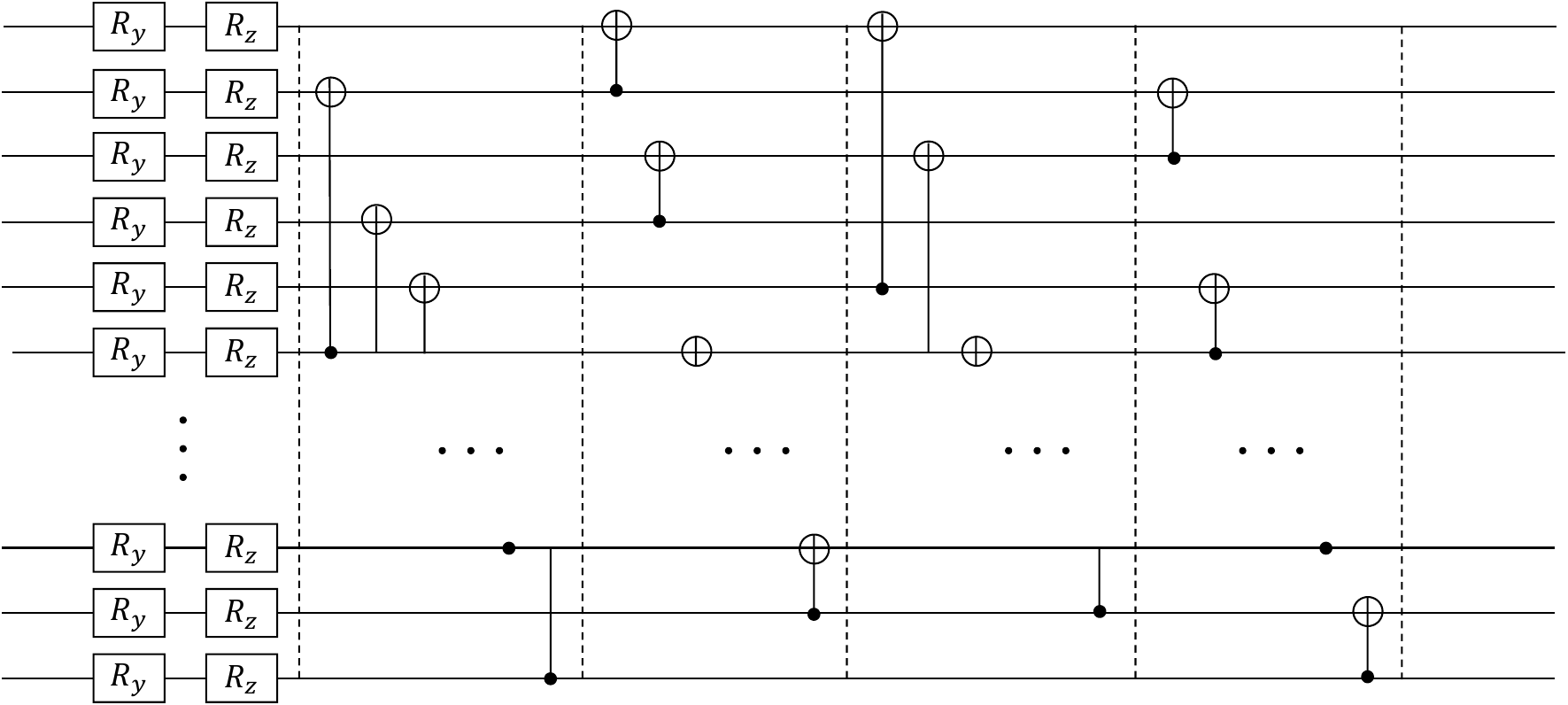}
\caption{\label{fig:circ4}One layer of the 16-qubit 2D pairwise circuit. 
From left to right, the four CNOT groups correspond to the $U_1$, $L_1$, $U_2$, and $L_2$ sublayers defined in Fig.~\ref{fig:pair2D}.}
\end{center}
\end{figure*}

This design covers all horizontal and vertical nearest-neighbor edges of the $4\times4$ grid without introducing nonlocal gates. 
Each of the four sublayers contains six non-overlapping CNOT gates that can be executed in parallel on hardware with the same nearest-neighbor connectivity.
Overall, this 2D pairwise topology provides a compact and hardware-compatible mechanism for enhancing entanglement propagation in variational quantum algorithms~\cite{McClean:2015vup, Cerezo:2020jpv}, QML models~\cite{Schuld_2014}, and other PQC  structures defined on 2D qubit arrays.
The quantum circuit implementation of the 2D pairwise ansatz is shown in Fig.~\ref{fig:circ4}.

\section{\label{sec:3} Evaluation Metrics}

The performance of  PQCs is determined not only by the range of quantum states they can represent but also by the trainability of their variational parameters. 
Therefore, when comparing different ansatz  structures, it is essential to evaluate both their expressibility and trainability.

Expressibility reflects the capability of PQCs to explore and cover the Hilbert space through the quantum states or unitary transformations they can generate. 
In general, a more expressive ansatz can approximate a broader class of target quantum states. 
However, excessive expressibility may also lead to optimization difficulties. 
Trainability, on the other hand, characterizes the ease with which the variational parameters of PQCs can be optimized and is therefore closely related to the practical performance of variational quantum algorithms.
Consequently, the design and analysis of PQCs require a careful balance between expressibility and trainability.

To systematically evaluate the performance of different initial states ansatze, we use three complementary metrics calculated from state vectors simulations in \verb"Qiskit"~\cite{qiskit2024} to evaluate all four ansatze. 
The KL divergence and the $t=2$ frame potential are adopted to characterize expressibility, while the gradient variance is used as a measure of trainability.
Together, these metrics provide complementary measures of expressibility and trainability, enabling a comprehensive assessment of PQC performance.
The KL divergence and the $t=2$ frame potential characterize complementary aspects of expressibility. 
Gradient variance is used only as a trainability proxy for the specified observable, initialization distribution, and parameter-selection rule. 
It does not by itself measure optimizer performance or establish barren-plateau scaling.

\subsection{\label{sec:3.1}Using KL divergence to evaluate expressibility}

The KL divergence quantifies the discrepancy between the fidelity distribution generated by an ansatz and the corresponding fidelity distribution of Haar-random states~\cite{Sim_2019,Kullback:1951zyt}.
To construct these probability distributions, the fidelity between two independently generated output states is defined as
\begin{equation}
F_{a,b}=|\langle\psi^{({a})}|\psi^{({b})}\rangle|^{2},
\tag{7}
\label{eq.3.1}
\end{equation}
where $\langle\psi^{({a})}|$ and $\langle\psi^{({b})}|$ correspond to the output states, respectively. 
The fidelity satisfies $F_{a,b}\in [0, 1]$, with $F_{a,b}=1$ if and only if $\langle\psi^{({a})}|=\langle\psi^{({b})}|$
up to a global phase, and $F_{a,b}=0$ if and only if the states are orthogonal.
For a $d$-dimensional Hilbert space, the fidelity between two independent Haar-random states has the probability density function~\cite{Sim_2019}
\begin{equation}
P_{Haar}(F)=(d-1)(1-F)^{d-2},\;\;F\in[0,1],
\tag{7}
\label{eq.3.2}
\end{equation}
where $d=2^{N}$, it denotes the dimension of the Hilbert space.
The corresponding cumulative distribution function
(CDF) is
\begin{equation}
C_{Haar}(f)=Pr(F\leq f)=1-(1-f)^{d-1},
\tag{8}
\label{eq.3.3}
\end{equation}
for $d \gg 1$, this distribution is sharply peaked near $F = 0$, indicating that two independent Haar-random states are nearly orthogonal with high probability.

To compare the empirical fidelity distribution to $P_{Haar}$, we divide the interval $[0, 1]$ into $B$ bins using quantile boundaries derived from the Haar CDF. 
Specifically, we choose $B+1$ bin edges $\{f_{k}\}^{B}_{k=0}$ such that each bin $[f_{k-1}, f_{k})$ has equal Haar probability $1/B$:
\begin{equation}
f_{k}=1-(1-\frac{k}{B})^{\frac{1}{d-1}},\;\; k=0,1,...,B,
\tag{9}
\label{eq.3.4}
\end{equation}
by construction, $f_0 = 0$ and $f_B = 1$.

To construct the empirical fidelity distribution, we independently sample two parameter vectors $\boldsymbol{\theta}^{(a)}_m$ and $\boldsymbol{\theta}^{(b)}_m$ for each $m=1,\ldots,M$, generate the corresponding output states, and calculate $F_m=|\langle\psi(\boldsymbol{\theta}^{(a)}_m)|\psi(\boldsymbol{\theta}^{(b)}_m)\rangle|^2$. 
The resulting values $\{F_m\}_{m=1}^{M}$ are assigned to the $B$ bins defined above.

Let $N_k$ denote the number of sample fidelity values falling within the $k$-th interval of the above partition, such that $\sum_{k=1}^B N_k = M$.
The empirical probability in bin $k$ is
\begin{equation}
\hat{p}_k=\frac{N_k}{M}.
\tag{10}
\label{eq.3.5}
\end{equation}
Based on the quantile construction, the reference probability in the $k$-th interval under the distribution of Haar-random state is
\begin{equation}
{q_k}=\int_{f_{k-1}}^{f_k} P_{Haar}(F)\,dF=\frac{1}{B}.
\tag{11}
\label{eq.3.6}
\end{equation}
The KL is defined as
\begin{equation}
D_{KL}=\sum_{k=1}^{B} \hat{p}_k\log\frac{\hat{p}_k}{q_k},
\tag{12}
\label{eq.3.7}
\end{equation}
Substituting Eqs.~(\ref{eq.3.5}) and (\ref{eq.3.6}) into Eq.~(\ref{eq.3.7}) yields the KL divergence,
\begin{equation}
D_{KL}=\sum_{k=1}^{B} \hat{p}_k\log\frac{\hat{p}_k}{q_k}=\sum_{k=1}^{B}\frac{N_k}{M}\log\frac{N_k/M}{1/B},
\tag{13}
\label{eq.3.7}
\end{equation}
this quantity is non-negative and vanishes if and only if
$\hat{p}_k=q_k$ for all $k$. 
In the numerical implementation, bins with $N_k=0$ contribute zero to the KL divergence, following the convention $0\log 0=0$.

To estimate the statistical uncertainty, $N_{\mathrm{boot}}$ bootstrap resamples are performed. 
For the $j$-th resampling, $M$ fidelities are drawn with replacement from the original sample ${F_m}$, the corresponding probabilities ${\hat{p}_k^{(j)}}$ are recomputed, and the KL divergence $D_{KL}^{(j)}$ is evaluated. 
The reported standard error is given by
\begin{equation}
\sigma_{D_{KL}}=\sqrt{\frac{1}{N_{boot}}\sum_{j=1}^{N_{boot}}(D_{KL}^{({j})}-\bar{D}_{KL})^2},
\tag{14}
\label{eq.3.8}
\end{equation}
where the mean value $\bar{D}_{KL}=\frac{1}{N_{\mathrm{boot}}}\sum_{j=1}^{N_{\mathrm{boot}}}D_{KL}^{(j)}$.

To summarize, the steps to evaluate the expressibility of ansatze using the KL divergence are listed as follows,
\begin{enumerate}
\item Sampling $M$ independent parameter pairs $\{\boldsymbol{\theta}^{(a)}_m,\boldsymbol{\theta}^{({b})}_m\}^{M}_{m=1}$ and computing their fidelities $F_m$.
\item Predetermining the $B$ bin boundaries $\{{f_k}\}_{k=0}^{B}$. 
Each fidelity $F_m$ is assigned to the bin satisfying $f_{k-1}\leq F_m < f_k$, and the resulting bin counts $N_k$ are computed.
\item Calculating the $D_{KL}$ divergence based on the empirical probability and the reference probability.
\item Sampling with replacement from $M$ fidelities, recalculate $\hat{p}_k^{(b)}$ and $D_{KL}^{(b)}$, repeat this $N_{boot}$ times, and calculate the standard deviation.
\end{enumerate}

\subsection{\label{sec:3.3}Using t=2 frame potential to evaluate expressibility} 

The frame potential provides a rigorous measure of how closely the ensemble generated by an ansatz approximates the Haar-random ensemble.
In particular, the $t=2$ frame potential quantifies the proximity of the generated ensemble to a state $2$-design~\cite{Renes_2004, Dankert_2009}.
Unlike the KL divergence, which measures the discrepancy between the fidelity distribution generated by an ansatz and the corresponding Haar-random fidelity distribution, the frame potential probes the second moment of the distribution of state overlaps and therefore provides complementary information about expressibility.
For $N$ qubits, with $d=2^N$, its theoretical lower bound is $2/[d(d+1)]$. 
A value closer to this lower bound indicates closer agreement with the reference second moment and therefore higher expressibility.

To construct the frame potential, consider two independently sampled parameter vectors $\boldsymbol{\theta}^{(a)}$ and $\boldsymbol{\theta}^{(b)}$.
The corresponding output states are
$|\psi^{(a)}\rangle=U(\boldsymbol{\theta}^{(a)})
|0\rangle^{\otimes N}$ and 
$|\psi^{(b)}\rangle=U(\boldsymbol{\theta}^{(b)})
|0\rangle^{\otimes N}$,
the squared overlap between these two states is defined as
\begin{equation}
O_{ab}=\left|\langle\psi^{(a)}|\psi^{(b)}\rangle
\right|^2,
\tag{15}
\label{eq:3.10}
\end{equation}
note that  $O_{ab}\in[0,1]$.
For $t=2$, the frame potential is defined as
\begin{equation}
F^{(2)}
=
\mathbb{E}_{\boldsymbol{\theta}^{(a)},\boldsymbol{\theta}^{(b)}}
\left[
O_{ab}^{2}
\right],
\tag{16}
\label{eq:3.11}
\end{equation}
where the expectation value is taken over independently sampled parameter vectors.
For the $d$-dimensional Hilbert space, the corresponding Haar value is
\begin{equation}
F_{\mathrm{Haar}}^{(2)}
=
\frac{2}{d(d+1)},
\qquad
d=2^N,
\tag{17}
\label{eq:3.12}
\end{equation}
this quantity represents the minimum possible value of the frame potential. 
Therefore, a smaller frame potential indicates that the generated ensemble is closer to the Haar ensemble and exhibits higher expressibility.

In the numerical implementation, $M$ independent output states are generated. 
The frame potential is estimated using off-diagonal estimator
\begin{equation}
\hat{F}^{(2)}
=
\frac{1}{M(M-1)}
\sum_{\substack{a,b=1\\a\neq b}}^{M}
O_{ab}^{2}
\tag{18}
\label{eq:3.13}
\end{equation}
The diagonal terms $O_{aa}^{2}=1$ are excluded to avoid finite-sample bias.
The off-diagonal overlap values are not independent because each sampled state contributes to multiple state pairs, and the pairs $(a,b)$ and $(b,a)$ give identical values. 
We therefore estimate the uncertainty using a delete-one-state jackknife, i.e., each jackknife replicate removes one sampled state together with all overlap pairs involving that state and then recomputes $\hat{F}^{(2)}$. 

The numerical implementation estimates the uncertainty with a delete-one-state jackknife. For each $a=1,\ldots,M$, delete state $a$ and all overlap pairs containing it, and define the leave-one-out estimate,
\begin{equation}
\widehat F^{(2)}_{(-a)}=\frac{1}{(M-1)(M-2)}\sum_{\substack{b,c=1\\b\neq c,\;b\neq a,\;c\neq a}}^{M}O_{bc}^{2}.
\tag{19}
\label{eq:3.14}
\end{equation}
The resulting jackknife standard error is,
\begin{equation}
\sigma_{\widehat F^{(2)}}=\sqrt{\frac{M-1}{M}\sum_{a=1}^{M}\left(\widehat F^{(2)}_{(-a)}-\frac{1}{M}\sum_{c=1}^{M}\widehat F^{(2)}_{(-c)}\right)^2}.
\tag{20}
\label{eq:3.15}
\end{equation}
Thus, the uncertainty calculation removes one sampled state at a time and does not bootstrap the correlated list of pairwise overlaps.

The procedure for evaluating expressibility using the frame potential can be summarized as follows:
\begin{enumerate}
\item Sample $M$ independent parameter vectors and generate the corresponding output states.
\item Compute the squared overlaps $O_{ab}$ between all distinct pairs of output states.
\item Evaluate the frame potential estimator $\hat{F}^{(2)}$ using Eq.~(\ref{eq:3.11}).
\item Compute the Haar value $F_{\mathrm{Haar}}^{(2)}$ using Eq.~(\ref{eq:3.12}).
\item Assess expressibility by comparing $\hat{F}^{(2)}$ with $F_{\mathrm{Haar}}^{(2)}$. 
A closer agreement indicates stronger expressibility.
\item Estimate the statistical uncertainty using the delete-one-state jackknife defined above.
\end{enumerate}

\subsection{\label{sec:3.4}Gradient Variance as a Trainability Proxy} 

Gradient variance is a widely used metric for assessing the trainability of parametric quantum circuits.
Unlike the KL divergence and the frame potential, which characterize expressibility, the gradient variance serves as a proxy for trainability by quantifying the strength of the optimization signal. 
A larger gradient variance generally corresponds to more informative gradients and improved trainability, whereas an exponentially vanishing gradient variance indicates the onset of barren plateau behavior.

We define the gradient variance using the expectation value of a string of Pauli-$Z$ matrices,
\begin{equation}
E(\boldsymbol{\theta})=\langle\psi(\boldsymbol{\theta})|\hat Z_{\mathrm{global}}|\psi(\boldsymbol{\theta})
\rangle=\sum_{x=0}^{d-1}(-1)^{w(x)}|c_x(\boldsymbol{\theta})|^2,
\tag{20}
\label{eq:3.16}
\end{equation}
where $\hat Z_{\mathrm{global}}$ is the global observable,
$\hat Z_{\mathrm{global}}=Z_0\otimes Z_1\otimes\cdots\otimes Z_{N-1}=\bigotimes_{k=0}^{N-1}Z_k$, and $Z_k$ is the Pauli-$Z$ operator acting on qubit $k$.
$w(x)$ is the Hamming weight, which denotes the number of $1$ in the binary representation of the integer $x$, and
$c_x(\boldsymbol{\theta}) = \langle 
x|\psi(\boldsymbol{\theta})\rangle$ is the complex amplitude of computational basis state $|x\rangle$ in the state $|\psi(\boldsymbol{\theta})\rangle$.	

A total of $M$ independent random initialization vectors
$\{\boldsymbol{\theta}^{(m)}\}_{m=1}^{M}$
are sampled, with each parameter drawn uniformly from the interval $[0,2\pi)$.
For every initialization and every ansatz, the gradient is evaluated with respect to the same parameter, $\theta_0$, corresponding to the first $R_y$ gate on qubit 0 in the first rotation layer,
\begin{equation}
g_m=\left.\frac{\partial E}{\partial\theta_0}\right|_{\boldsymbol{\theta}^{(m)}},
\tag{22 }
\label{eq:3.17}
\end{equation}
the partial derivative of the expectation value with respect to $\theta_i$ can be evaluated exactly on a quantum computer via the parameter-shift rule~\cite{Schuld:2018aiz},
\begin{equation}
\frac{\partial E}{\partial \theta_i}
=\frac{E(\theta_i+\pi/2)-E(\theta_i-\pi/2)}{2},
\tag{22}
\label{eq:3.18}
\end{equation}
where $E(\theta_i \pm \pi/2)$ denotes the expectation value
$E(\boldsymbol{\theta}')$ evaluated at the parameter vector
$\boldsymbol{\theta}'$, which is identical to
$\boldsymbol{\theta}$ except that the $i$-th component is shifted by $\pm \pi/2$.
We fix $i=0$ for every initialization and every ansatz, where $\theta_0$ parameterizes the first $R_y$ gate on qubit $0$ in the first rotation layer. 
We sample $M=10{,}000$ independent parameter vectors and evaluate $g_m=\left.\partial E/\partial\theta_0\right|_{\boldsymbol{\theta}^{(m)}}$ using the parameter-shift rule. 

The trainability of the ansatz is characterized by the variance of this gradient over randomly initialized parameters,
\begin{equation}
\mathrm{Var}(g)
=\frac{1}{M}\sum_{m=1}^Mg_m^2-(\frac{1}{M}\sum_{m=1}^Mg_m)^2.
\tag{23}
\label{eq:3.19}
\end{equation}
A larger variance indicates a larger typical gradient signal for this specified observable, parameter, and initialization distribution. 
It does not by itself determine optimization performance for other objective functions or parameter locations.
Barren-plateau behavior is identified through asymptotic scaling, typically an exponential suppression of gradient variance with the number of qubits for a specified cost and initialization ensemble. 
A depth scan at a single system size can reveal depth-dependent concentration, but cannot by itself establish this scaling.

To estimate the statistical uncertainty, $N_{\mathrm{boot}}$ bootstrap resamples are performed.
For the $j$-th resampling, we draw $M$ gradients with replacement from $\{g_m\}^M_{m=1}$, recompute the variance $Var^{(j)}[g]$,
The reported standard error is
\begin{equation}
\sigma_{Var}=\sqrt{\frac{1}{N_{boot}}\sum_{j=1}^{N_{boot}}({Var}^{(j)}[g]-\overline{\mathrm Var}[g])^2}
\tag{24}
\label{eq:3.20}
\end{equation}
where $\overline{\mathrm Var}[g]$ is the mean over bootstrap resamples.

The procedure for evaluating trainability using gradient variance can be summarized as follows:
\begin{enumerate}
\item Sample $M$ independent parameter vectors $\boldsymbol{\theta}^{(m)}$.
\item Compute the gradients $g_m=\partial E/\partial\theta_i$ using the parameter-shift rule.
\item Estimate the gradient variance using Eq.~$\ref{eq:3.19}$.
\item Assess trainability by comparing the resulting variance across different ansatze. 
For this fixed observable and parameter, a larger gradient variance indicates a larger typical gradient signal.
\item Perform $N_{\mathrm{boot}}$ bootstrap resamples and estimate the statistical uncertainty using Eq.~(\ref{eq:3.20}).
\end{enumerate}

\section{\label{sec:4}Numerical results}

All calculations use $N=16$ qubits and ansatz-layer counts $L=1,\ldots,6$. 
The KL-divergence calculation uses 10,000 independent parameter pairs, $50$ equal-reference-probability bins, and $20$ bootstrap resamples.
The frame-potential calculation uses 2,000 independently sampled states and delete-one-state jackknife errors. 
The fixed-parameter derivative calculation uses 10,000 independent initializations and 200 bootstrap resamples. 
Thus, $L$ is the repeated-layer count in every table and figure, while the uncertainty procedure is stated separately for each metric.

\subsection{\label{sec:4.1}KL divergence} 

\begin{table*}[htbp]
\centering
\begin{tabular}{c|c|c|c|c|c|c} 
\hline
Ansatz& $ L=1$ & $L=2$ & $ L=3$  & $ L=4$ & $ L=5$& $ L=6$ \\ 
\hline
1D\;pairwise&$0.5097$&$0.0706$&$0.0071$&$0.0022$&$0.0026$&$0.0022$\\
\hline
Circular &$0.2916$&$0.0237$&$0.0042$&$0.0025$&$0.0027$&$0.0025$\\
\hline
SCA &$0.2916$&$0.0313$&$0.0041$&$0.0019$&$0.0021$&$0.0019$\\
\hline
2D\;pairwise&$0.1092$&$0.0041$&$0.0033$&$0.0027$&$0.0032$&$0.0029$\\
\hline
\end{tabular}
\caption{The values of the KL divergence for different depths $L$.}
\label{table:kl}
\end{table*}
As a measure of expressibility, the KL divergence quantifies the discrepancy between the fidelity distribution generated by a random quantum circuit and the theoretical fidelity distribution of Haar-random states. 
A smaller KL divergence indicates that the state ensemble produced by the ansatz is closer to the Haar distribution, and therefore, exhibits higher expressibility.
The samples of $M = 10,000$ independent parameter pairs are selected for each ansatz, using $B = 50$ quantile bins.

\begin{figure}[htbp]
\begin{center}
\includegraphics[width=0.99\hsize]{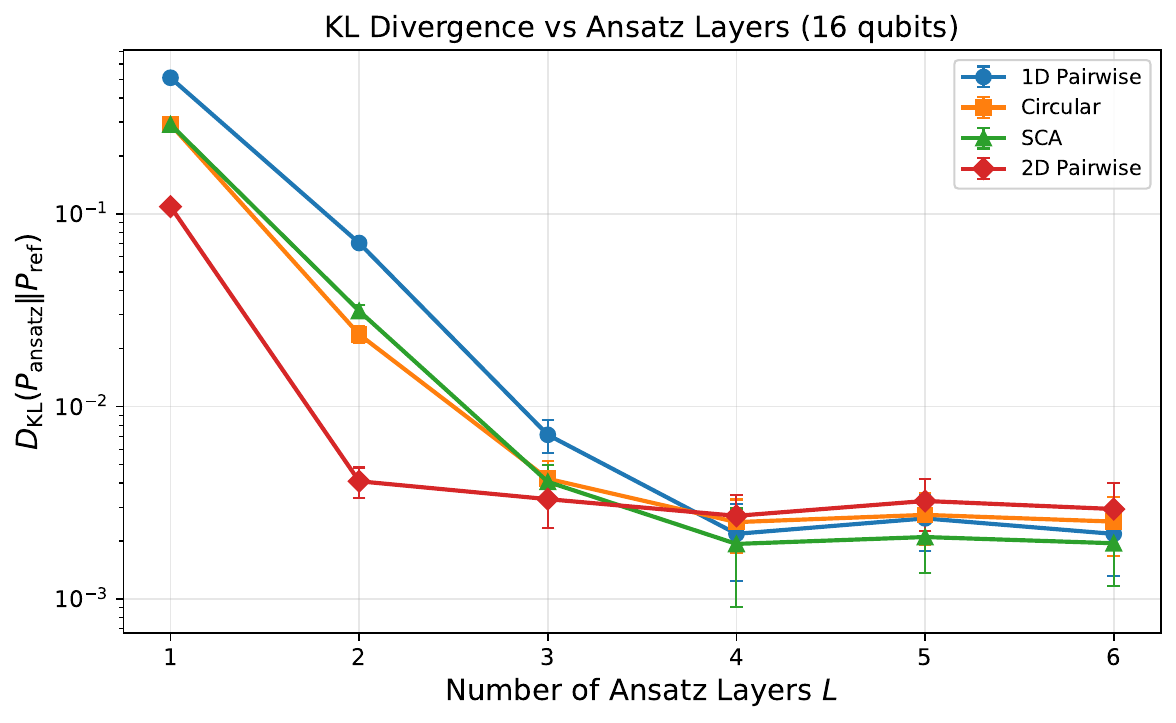}
\caption{\label{fig:kld}KL divergence between the sampled fidelity distribution and the uniformly random pure-state reference distribution as a function of the number $L$ of ansatz layers. 
Smaller values indicate closer agreement with the reference distribution. 
Error bars are bootstrap standard errors from 20 resamples of the 10,000 independent fidelities.}
\end{center}
\end{figure}

The numerical results are summarized in Table~\ref{table:kl} and Fig.~\ref{fig:kld}.
The error bars represent the standard errors estimated from bootstrap resampling.
As shown in Fig.~\ref{fig:kld}, the 2D pairwise ansatz achieves substantially lower KL divergence values than all 1D ansatze at shallow depths $L=1\sim2$, indicating that it attains a higher level of expressibility with fewer layers. 
As the circuit depth increases, the KL divergence decreases rapidly for all ansatze. 
At $L=4$--$6$, SCA has slightly lower central KL-divergence values than Circular and 1D pairwise, but the differences are smaller than the reported statistical uncertainties and do not establish an expressibility ranking.
Notably, all ansatze converge to similarly low KL divergence values beyond $L=3\sim4$, indicating that their generated state ensembles have become sufficiently close to the Haar distribution.
Consequently, further increases in circuit depth yield only limited improvements in expressibility.

\subsection{\label{sec:4.2}Frame potential} 

\begin{table*}[htbp]
\centering
\begin{tabular}{c|c|c|c|c|c|c|c} 
\hline
Ansatz& $ L=1$ & $L=2$ & $ L=3$  & $ L=4$ & $ L=5$& $ L=6$& Lower bound \\ 
\hline
1 D\;pairwise &$3.18 \times 10^{-9}$&$9.25\times 10^{-10}$&$5.32\times 10^{-10}$&$4.74\times 10^{-10}$&$4.66\times 10^{-10}$&$4.66\times 10^{-10}$&$4.66\times 10^{-10}$\\
\hline
Circular &$1.96\times 10^{-9}$&$6.50\times 10^{-10}$&$4.96\times 10^{-10}$&$4.71\times 10^{-10}$&$4.68\times 10^{-10}$&$4.65\times 10^{-10}$&$4.66\times 10^{-10}$\\
\hline
SCA&$1.96\times 10^{-9}$&$6.87\times 10^{-10}$&$5.08\times 10^{-10}$&$4.75\times 10^{-10}$&$4.67\times 10^{-10}$&$4.67\times 10^{-10}$&$4.66\times 10^{-10}$\\
\hline
2D\;pairwise&$1.16\times 10^{-9}$&$5.03\times 10^{-10}$&$4.70\times 10^{-10}$&$4.67\times 10^{-10}$&$4.65\times 10^{-10}$&$4.66\times 10^{-10}$&$4.66\times 10^{-10}$\\
\hline
\end{tabular}
\caption{Same as Table~\ref{table:kl}, but for $t=2$ frame potential.}
\label{table:fp}
\end{table*}

\begin{figure}[htbp]
\begin{center}
\includegraphics[width=0.99\hsize]{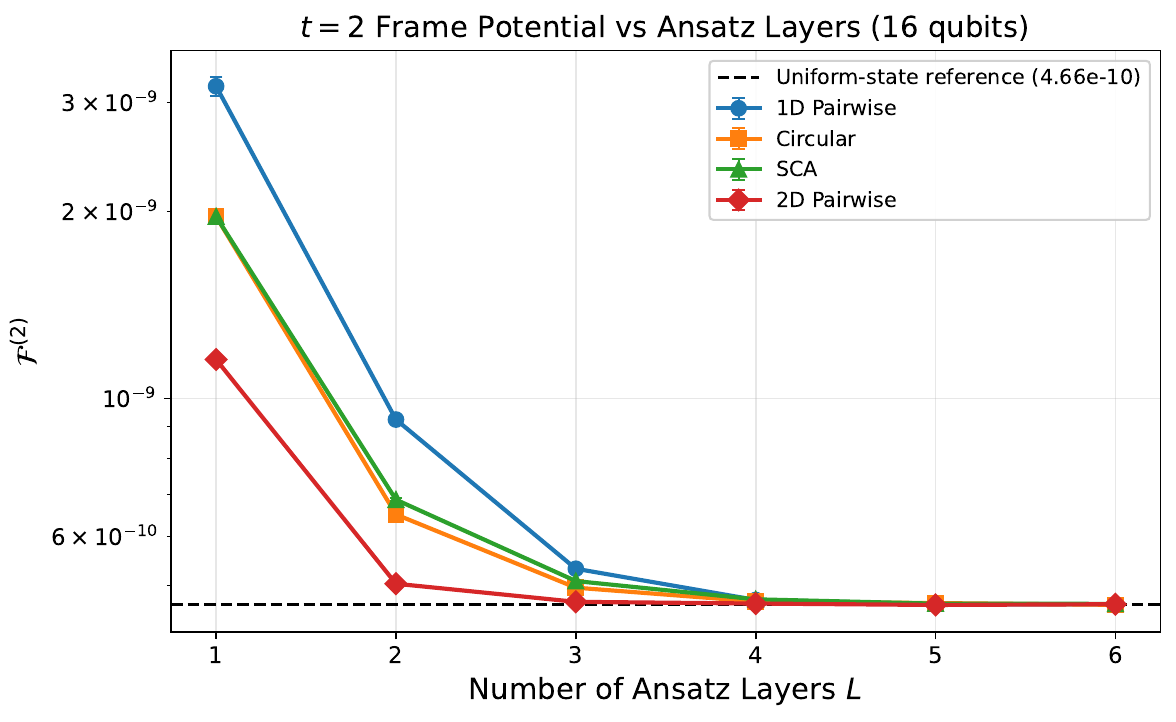}
\caption{\label{fig:fp}The $t=2$ frame potential as a function of the number $L$ of ansatz layers. 
For the fixed 16-qubit system, the dashed line marks the theoretical lower bound $2/[d(d+1)]$. 
Values closer to this line reproduce the reference second moment of state overlaps more accurately. 
Error bars are delete-one-state jackknife standard errors computed from the 2,000 independently sampled states.
}
\end{center}
\end{figure}

As a measure of expressibility, the frame potential quantifies the degree to which the state ensemble generated by an ansatz resembles that produced by Haar-random states. 
The theoretical frame potential is bounded below by $2/[d(d+1)]$. 
Because a finite-sample estimate can fluctuate slightly below this bound, agreement is assessed by the distance from the bound rather than by treating every smaller estimate as better.
We calculated this metric using the same parameters as for the KL divergence, but with $M = 2,000$ independent random states at each depth.
The numerical results are listed in Table~\ref{table:fp} and Fig.~\ref{fig:fp}.

Table~\ref{table:fp} and Fig.~\ref{fig:fp} show that the two-dimensional pairwise circuit is closest to the fixed 16-qubit theoretical lower bound at $L=1$.
The one-dimensional pairwise circuit approaches the reference value most slowly at small $L$. 
By $L=3$--$4$, all four central values are close to $2/[d(d+1)]$, meaning that their sampled fourth-power overlaps closely match the reference benchmark. 
Further layers therefore produce only small changes in this metric.

\subsection{\label{sec:4.3}Gradient variance } 

\begin{table*}[htbp]
\centering
\begin{tabular}{c|c|c|c|c|c|c} 
\hline
$Ansatz$& $ L=1$ & $L=2$ & $ L=3$  & $ L=4$ & $ L=5$& $ L=6$\\ 
\hline
$1D\;\mathrm{pairwise}$ &$1.97\times 10^{-5}$&$1.81\times 10^{-5}$&$9.79\times 10^{-6}$&$8.33\times 10^{-6}$&$7.79\times 10^{-6}$&$7.73\times 10^{-6}$\\
\hline
$\mathrm{Circular}$ &$1.30\times 10^{-5}$&$1.48\times 10^{-5}$&$1.06\times 10^{-5}$&$8.54\times 10^{-6}$&$8.05\times 10^{-6}$&$7.71\times 10^{-6}$\\
\hline
$\mathrm{SCA}$ &$1.30\times 10^{-5}$&$1.10\times 10^{-5}$&$9.22\times 10^{-6}$&$8.20\times 10^{-6}$&$7.79\times 10^{-6}$&$7.84\times 10^{-6}$\\
\hline
$2D\;\mathrm{pairwise}$ &$8.57\times 10^{-6}$&$7.62\times 10^{-6}$&$7.79\times 10^{-6}$&$7.81\times 10^{-6}$&$7.56\times 10^{-6}$&$7.59\times 10^{-6}$\\
\hline
\end{tabular}
\caption{Gradient variance of the fixed first-layer parameter $\theta_0$ for $M=10{,}000$ independent random initializations.}
\label{table:gv}
\end{table*}

\begin{figure}[htbp]
\begin{center}
\includegraphics[width=0.99\hsize]{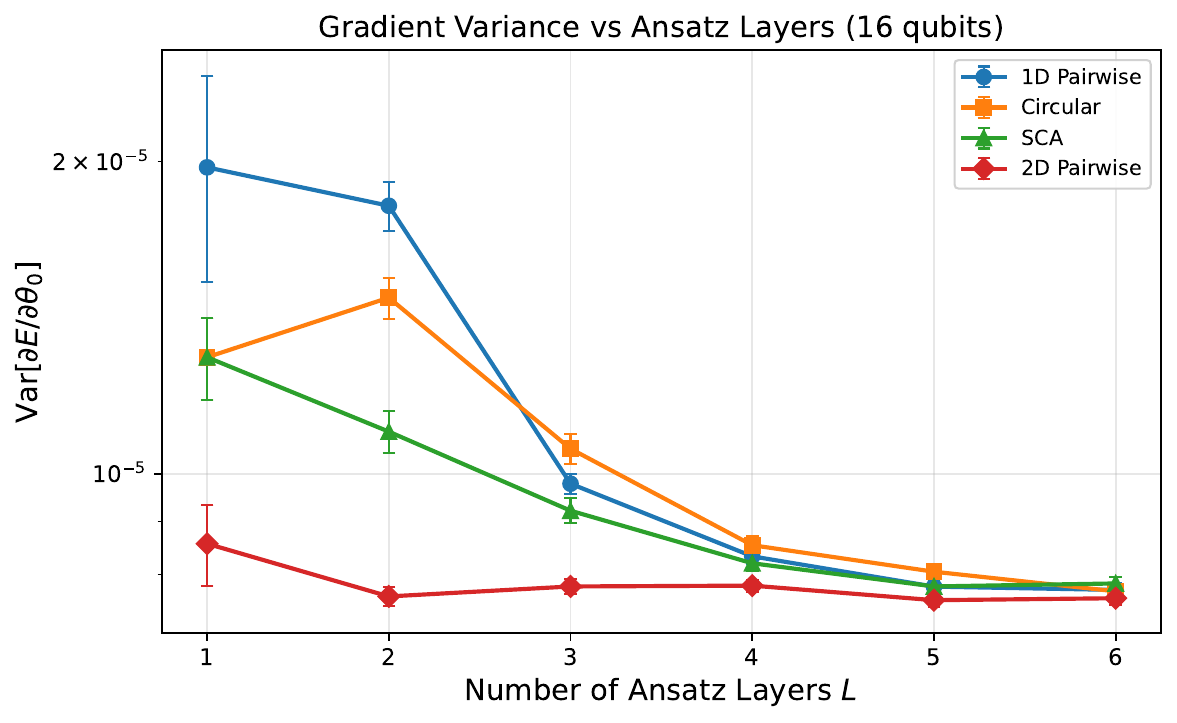}
\caption{\label{fig:gv}Variance of $\partial E/\partial\theta_0$ as a function of the number $L$ of ansatz layers, where $E=\langle\bigotimes_{k=0}^{15}Z_k\rangle$ and $\theta_0$ is the first $R_y$ angle on qubit 0. 
Each point uses 10,000 independent random initializations. 
Error bars are bootstrap standard errors from 200 resamples.}
\end{center}
\end{figure}

Table~\ref{table:gv} and Fig.~\ref{fig:gv} report the variance of the derivative of $\langle\bigotimes_{k=0}^{15}Z_k\rangle$ with respect to the fixed first-layer angle $\theta_0$. 
The two-dimensional circuit has the smallest central variance at every $L$. 
The bootstrap error bars separate it from all three one-dimensional circuits at $L=1$--$4$. At $L=5$, only the difference from Circular remains clearly resolved. 
At $L=6$, the four estimates overlap within their bootstrap uncertainties. 
The calculation establishes a smaller fixed-parameter gradient variance for the specified Pauli-$Z$ string at low and intermediate $L$.

\section{\label{sec:5}Summary}

In this work, four ansatze including 1D pairwise, Circular, SCA, and 2D pairwise are systematically compared using complementary metrics of expressibility and trainability.
At $L=1$ and $2$, the 2D pairwise ansatz has the smallest KL divergence, and its $t=2$ frame potential is closest to the fixed 16-qubit theoretical lower bound. Both measures therefore show the strongest shallow-layer expressibility for the 2D pairwise ansatz.
As the circuit depth increases, all ansatze approach similar levels of expressibility, with only minor differences remaining beyond $L\approx3-4$. 
In the intermediate layer range, the small differences among the central KL-divergence values are not resolved by the reported uncertainties.

For the specified Pauli-$Z$ string and the fixed first $R_y$ parameter, the 2D pairwise ansatz has the smallest gradient variance at low and intermediate layer counts. 
This result does not establish that the complete optimization landscape is flatter or that the 1D ansatze are generally more trainable. 
Because only $N=16$ is studied, the calculation does not test barren-plateau scaling with system size.
Overall, the layer-matched comparison shows that the 2D pairwise circuit approaches the Haar-based expressibility benchmarks more rapidly at shallow depth. 
For the expectation value of $Z_0\otimes\cdots\otimes Z_{15}$, the fixed-parameter gradient variance is smaller for the 2D circuit at $L=1$--$4$, while the differences narrow at $L=5$ and are statistically unresolved at $L=6$. 
This is a result for one observable, one parameter, and one system size. 
Establishing barren-plateau behavior or task-level trainability requires calculations at several values of $N$ and direct optimization with task-specific objective functions.

\section*{Data availability statement}
Both the data and the code that support the findings of this study are openly available at \url{https://www.modelscope.cn/datasets/nbalexis/Expressibility_and_Trainability_Data_for_2D_Pairwise_Ansatz}.

\begin{acknowledgements}
This work was supported in part by the National Natural Science Foundation of China under Grants Nos. 12575106 and 12147214, the Basic Research Projects of Universities in Liaoning Province (Grant No.~LJ212510165024).
\end{acknowledgements}

\bibliographystyle{apsrev4-2}
\bibliography{ansatz}

\end{document}